\documentclass[preprint,12pt]{elsarticle}
\usepackage{color}
\usepackage{amsmath}
\usepackage{amssymb}
\journal{Physica A}

\begin{document}

\begin{frontmatter}

\title{Small fraction of selective cooperators can elevate general wellbeing significantly}

\author[as]{Hsuan-Wei Lee}
\author[as]{Colin Cleveland}
\author[mfa]{and Attila Szolnoki}
\address[as]{Institute of Sociology, Academia Sinica, Taiwan}
\address[mfa]{Institute of Technical Physics and Materials Science, Centre for Energy Research, P.O. Box 49, H-1525 Budapest, Hungary}

\begin{abstract}
A cooperative player invests effort into a common venture without knowing the partner's intention in advance. But this strategy can be implemented in various ways when a player is involved in different games simultaneously. Interestingly, if cooperative players distinguish their neighbors and allocate all their external investments into the most successful partner's game exclusively then a significant cooperation level can be reached even at harsh circumstances where game parameters would dictate full defection otherwise. This positive impact, however, can also be reached when just a smaller fraction of players apply this sophisticated investment protocol during the game. To confirm this hypothesis we have checked several distributions that determine the fraction of supporting players who apply the mentioned selective investment protocol. Notably, when these players are not isolated, but their influences percolate then the whole population may enjoy the benefit of full cooperation already at a relatively low value of the synergy factor which represents the dilemma strength in the applied public goods game.
\end{abstract}

\begin{keyword}
\texttt social dilemma\sep selective cooperation \sep strategy-neutral mechanism
\end{keyword}

\end{frontmatter}

\section{Introduction}
\label{intro}

To understand and explain cooperation among self-interested competitors is a long-standing intellectual task because an evident argument would dictate the opposite behavior during individual decisions \cite{nowak_sa95,sigmund_10,nowak_11}. This temptation to exploit others makes the problem especially important and the related dilemma stimulates scientific activities to find mechanisms to avoid it \cite{nowak_s06,szolnoki_prl12,hauert_s02,chen_xj_pa08,jimenez_jtb08,yang_r_epjb20,zhang_bj_pa19,zhong_lx_csf20}. In the last three decades several reasonable and less plausible circumstances were detected as a decisive tool to maintain cooperative acts and avoid the spreading of defection \cite{perc_pr17,szabo_bcp08,ohdaira_srep17,danku_epl18,chen_q_csf16,amaral_pre18,deng_c_epl18,cardinot_njp19,li_b_pa19,li_xp_pla20,wang_l_amc21}. Since the interaction of system members can be frequently described by the competition of the mentioned strategies therefore related results and observations can expect the interest of several seemingly separated disciplines, including biology, ecology, economy or social sciences \cite{cremer_srep12,bazeia_srep17,chica_srep19,ball_n07,allen_pre18,wang_z_pre14}.

In some cases it seems reasonable to punish defectors who just exploit others' efforts without contributing to the common pool, or rewarding those who do \cite{szolnoki_pre11,takesue_epl18,gao_sp_pla20b,cheng_f_amc20,szolnoki_epl10}. Furthermore, it also seems plausible to cooperate with those who have a higher reputation because of their previous positive acts and break interaction with those who have a negative fame \cite{fehr_n04,chen_xj_pone12b,yang_hx_pa19}. In general, the mentioned tools require an extra effort from some competitors by bearing the cost of punishment/reward or demand additional cognitive skill from players to record and keep in mind previous interactions. Evidently, these circumstances reduce the applicability of a general model, hence provide a limited solution to the above exposed general problem.

In game theory the term payoff, as the result of interactions with other competitors, offers a general concept to rank the success of competitors during the evolution process \cite{maynard_82}. Its higher value could be a clear sign of which player is worth imitating among the available partners for a better performance. But its value can be also used for additional purpose even if we still keep the traditional way of learning protocol and assume that an alternative strategy of a model partner becomes more attractive depending on the payoff difference of the related players \cite{szabo_pre98}. Think about a public goods game ({\it PGG}) in a spatially structured population where every player is involved in several games: in one, where she is the focal player and in those which are organized by neighboring $(G-1)$ partners \cite{szolnoki_pre09c}. According to the traditional protocol, a cooperator player contributes a fixed amount of $c$ to every game where she is involved and the sum of contributions is enhanced and redistributed among all group members independently on their inputs. But a cooperator may think differently to avoid the waste of positive acts and to ensure the best return of her investments. While she still invests the total $(G-1) \cdot c$ amount to external pools, but she does it in a highly selective way: she concentrates all the mentioned amount into a single pool that is organized by the most successful neighbor, hence the remaining $G-2$ groups enjoy nothing from the cooperative act. This attitude, called as ``selective cooperator'' is proved to be an efficient way to reach a high cooperation level globally \cite{szolnoki_amc20}. If every player follows this investment protocol then cooperators can survive even at harsh external conditions, which are characterized by the low value of the enhancement or synergy factor. 

For the sake of completeness, we note that several models have been suggested in the last decade where cooperators invested into neighboring groups in a non-uniform way \cite{cao_xb_pa10,huang_k_pone15}. But in general, these works utilized something special feature of the model setup. In some cases players who are organizing a larger group may expect higher investments from neighbors \cite{cao_xb_pa10,wang_hc_pa18}, or in some cases reputation, based on previous positive acts of the actual player, served as a compass where to invest more \cite{yang_hx_pa19}. Toward more sophisticated strategies, this research path can be extended by investment protocols which demand even higher cognitive skill from players to record previous benefits from the actual pool \cite{vukov_jtb11}. However, such an extra effort seems plausible to be considered via an extra cost for those who apply it. One may claim that for similar reason the above described success-based selective investment protocol also assumes that players have extra knowledge about the success of their neighbors, hence additional cost might also be established. But this is not the case, because the mentioned information about the success of neighbors is already used implicitly during the broadly applied  microscopic step of strategy adoption protocol \cite{szabo_pre98}, hence we demand nothing extra from players comparing to the traditional model setup. 

Most importantly, in our present work we assume that we have a heterogeneous population where some players follow the traditional way and invest in all available pools uniformly and there are others who follow the selective investment protocol when they are in a cooperator state. We will show that there is no need for every player to be in selective mode, but a full cooperator state can still be reached. Hence, general well-being can be obtained already if a smaller portion of players follows this more sophisticated, and perhaps less democratic, but effective, investment protocol. 

\section{Traditional and alternative investment protocols applied by selective players}
\label{def}

\begin{figure}
\centering
\includegraphics[width=5.5cm]{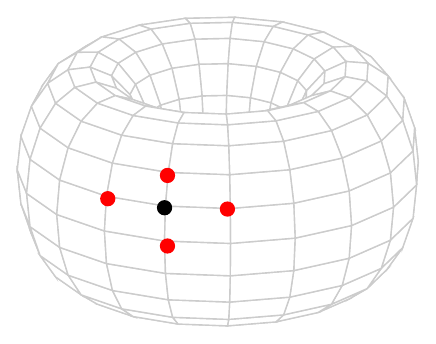}
\caption{Schematic presentation of the spatial relationship of the players on a torus structure. The black node with her four red neighbors are forming a group where they play a public goods game. The focal player, however, also participates in other $5$-member games organized by the four red nodes. Hence, similarly to all other players, the mentioned black node is involved in altogether $G=5$ games.}
\label{torus}
\end{figure}

By following the rules of the traditional $PGG$, the members of a group have two options, to contribute or not to a common pool. All contributions are summarized and multiplied by a synergy factor $r$. After the enhanced amount is distributed equally among all group members independently of their strategies. In our spatial population the public goods game is staged on a square lattice where periodic boundary condition is applied \cite{perc_jrsif13,wu_t_njp18}.  According to the interaction topology, the players are distributed on the surface of a torus shape. Figure~\ref{torus} describes a focal player, marked by black node, who organizes a game with four neighboring partners who are marked by red nodes. In agreement to the traditional setup the black player is also involved in the games organized by the neighbors, hence altogether every player participates in $G=5$ different games.
 
Importantly, we assume a regular interaction network because irregular graph would already ensure an obvious benefit for those nodes who have higher degree, hence larger personal income is expected to them \cite{cao_xb_pa10,santos_n08}. Since the presence of heterogeneity is a broadly recognized cooperator supporting condition \cite{szolnoki_epl07,perc_pre08,rong_zh_c19,amaral_rspa20,cheng_f_pa19,takeshue_epl19,pei_zh_njp17,zhu_p_epjb21,szolnoki_pre10b} therefore we want to avoid this effect and focus on the proposed investment protocol only.
  
Our extended model assumes that two types of players are present. There are traditional players who follow the standard investment protocol when they are in cooperator state. In this case they invest $c=1$ amount to each game where they are involved. Evidently, when they are in defector state then they contribute nothing to the common pool. Furthermore, there are also so-called ``selective players" who follow different investment policy. Such a player still contributes $c=1$ amount to her own group where she is the focal player. However, the remaining $(G-1)\cdot c$ resources are not distributed uniformly among the external groups where she is also involved. Instead, she invests this amount into a single game exclusively. Importantly, the external group which is awarded by the mentioned concentrated contribution is the one whose organizer has the highest payoff among the four neighboring partners. As a technical note, if there are more than one neighbors who have the highest payoff value then one of them is selected randomly to earn the mentioned contribution. Similarly to the previous case, when a selective player is in defector state then she just enjoys the contribution of other players without suffering from the cost of the game. At this point it is worth stressing that the suggested alternative investment protocol does not modify the average relation of defector and cooperator players because selector players altogether invest the same amount to games as normal cooperator players do.

Initially, each player on site $x$ is designated as a cooperator ($s_x = C$) or a defector ($s_x = D$) with equal probability, and each competitor is assigned to be a selective player with a probability $p$. Here $p$ is the outcome of a random variable (RV).

Importantly, the microscopic dynamics of a Monte Carlo (MC) procedure remains the same as in the traditional model. Thus in every elementary step a player $x$ and a neighboring $y$ player are selected randomly. During the strategy invasion process player $x$ changes its strategy from $s_x$ to $s_y$ with the probability
$$ W(s_x \to s_y) = \frac{1}{1 + \exp[(P_x - P_y)/K]}$$
where $P_x$ and $P_y$ are the accumulated payoff values of $x$ and $y$ originated from the $G$ games where they are involved. As usual, parameter $K$ determines the uncertainty of strategy adoption which was chosen to be $K=0.5$ for comparable results to the classic model \cite{szolnoki_pre09c}. A full MC step involves $L \times L$ elementary steps, hence on average every player has a chance to change her strategy. For the results presented below, we used lattices with the linear sides of $L = 100$ to $400$ where $10^5$ MC steps were generated to reach the stationary state for all parameter values. 

As noted, a selective cooperator contributes only to her own and to the most successful partner's games. Evidently, this act requires a preliminary knowledge about the latter player which is unavailable information in the first round. Therefore initially selective cooperators behave as a traditional cooperators without using the selective investment protocol. After the requested information become available and the system evolves into a state which is practically independent from the starting state.
After an initial relaxation period the system evolves into a stationary state where the time-averages of all strategies become constant. In this state we have measured the time-average of the portion of cooperator players which involves the traditional and selective cooperators as well.

In our experiments, we used different ways to designate selective players in the whole population. In case of deterministic distribution a fixed value of $p$ is used. When uniform distribution is applied then the random variable was used from the $(0, 2p$) interval for $p<1/2$ and from $(2p-1,1)$ interval for $1/2<p$ to ensure the same expected $p$ value as in the previous case. Our third distribution was a bimodal one where zero and 1 values were chosen appropriately to ensure the same average value as earlier. Evidently, for proper comparison the same expected value should be obtained for all distributions. Last, we also used truncated exponential and truncated power-law distributions to define the RV function.
 
As the range of probability is limited in $[0,1]$ interval, the exponential distribution and power-law distribution need to be truncated as their supports are in $[0,\infty)$. For the simplicity of calculation and program implementation, we write the truncated exponential distribution from the scratch and approach the truncated power-law distribution with Riemann approximation. In case of exponential distribution the probability is obtained via two steps. The first one is to generate the exponential distribution from a uniform distribution. The second one is truncate the distribution. For example suppose that $X$ is the exponential distribution. From the fact that quantile function of $X$ is $-\log(1-p)/\lambda$, we may generate $X$ with $p = $ uniform distribution in $[0,1]$. For the truncated, we may follow the forgetfulness property to set $X' = X - \lfloor X \rfloor$ and have the truncated distribution.

As for the power law distribution, the related function is $p(x) \approx 1/(x^r) $ with $r > 1$ and $x \in (0,\infty]$. For simplicity, we approximate the truncated power law distribution ``Riemannly" with partition distance $0.05$. Suppose $a$ is the desired expectation value for this RV. We use the bisection method to find the apropos $r$ with precision $10^{-5}$. Further details of the used distribution functions and the source code of simulations including the full instructions how to run it are available at \cite{source}.

\section{Results}
\label{results}

\begin{figure}
\centering
\includegraphics[width=12.5cm]{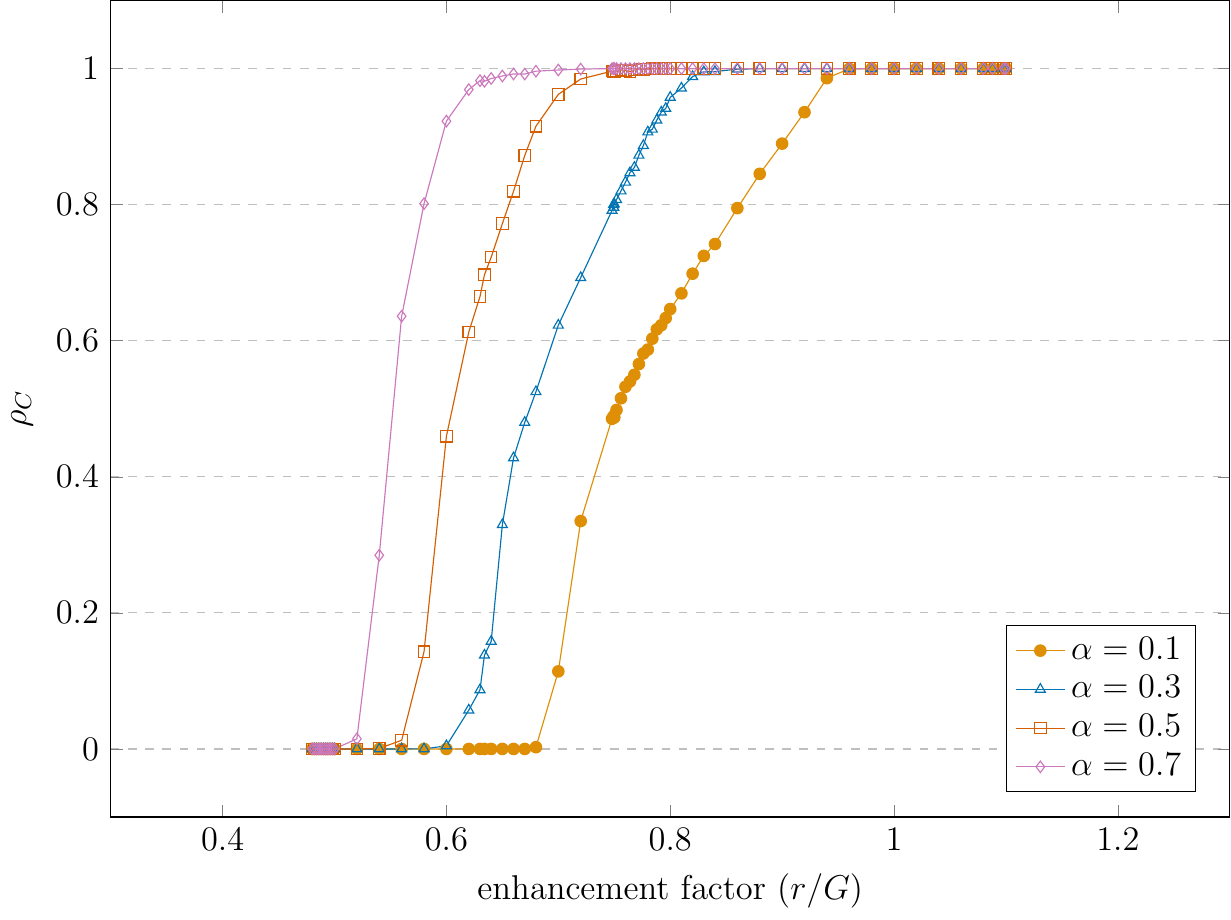}\\
\caption{Cooperation level in dependence of normalized synergy factor for bimodal distribution where $\alpha$ portion of players are in ``selective'' mode. The actual fraction of them is indicated in the legend. As we increase this portion, the tragedy of the commons state can be avoided for lower values of synergy factor.}
\label{bimodal_1}
\end{figure}

As the simplest case, we start our presentation with the results obtained for bimodal distribution when classes of players are clearly dichotomized. Set $\alpha=p$ to be the portion of the population who follow the selective investment protocol when they are cooperator while $1-\alpha$ fraction represents the traditional players who distribute their investment uniformly in case of cooperation. Evidently, both limit values of $\alpha$ restore a uniform population: $\alpha=0$ represents the traditional {\it PGG} model, while in the $\alpha=1$ limit we get back the uniform model discussed in Ref.~\cite{szolnoki_amc20}. Some representative results are shown in Fig.~\ref{bimodal_1} where the cooperation level is plotted according to the normalized synergy factor for different values of $\alpha$. These curves suggest that the cooperation level can be increased even if only just a portion of the population follows the selective investment protocol. Naturally the more the better principle is valid, hence we can reach a higher cooperation level for higher $\alpha$ values when the majority of players pay special attention to where to invest in the neighboring ventures.

\begin{figure}
\centering
\includegraphics[width=12.5cm]{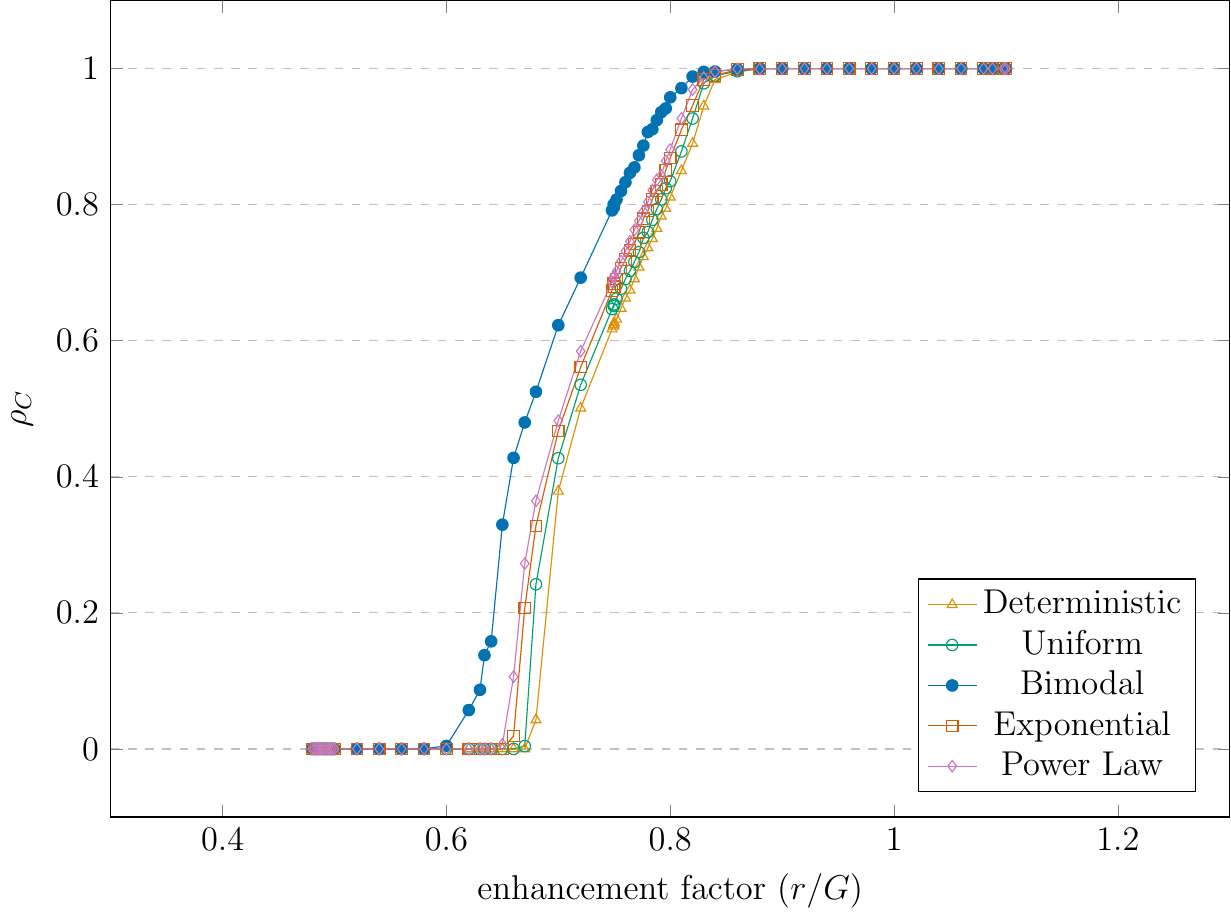}\\
\caption{Cooperation level in dependence of normalized synergy factor for various distributions where the expectation value of $\alpha$ is 0.3 for all cases. The applied functions are indicated in the legend. The comparison demonstrates that the presence of selective players is helpful for all cases.}\label{all}
\end{figure}

Next, we investigate several distributions that determine how individuals in the population apply traditional or selective investment protocol. In general, conceptually similar behavior can be observed as we illustrated in Fig.~\ref{bimodal_1}. Namely, as we increase the portion of those players who are in selective mode then the average cooperation level is increased. Because of the significantly different characteristics of the applied distributions, the direct comparison of results for various functions is difficult, but a common setting could be applied to compare the cooperation of those cases when the expectation levels of $\alpha$ are equal. The experimental result is shown Fig.~\ref{all}. It suggests that the observed effect is robust and there is no significant difference between the outcomes obtained by different distribution functions. Note that the bimodal distribution could provide the highest cooperative level in the end. In this case, to reach the same expectation level of $\alpha$, more individuals need to be in the selective mode and this gives a small benefit to this distribution.

\begin{figure}
\centering
\includegraphics[width=12.5cm]{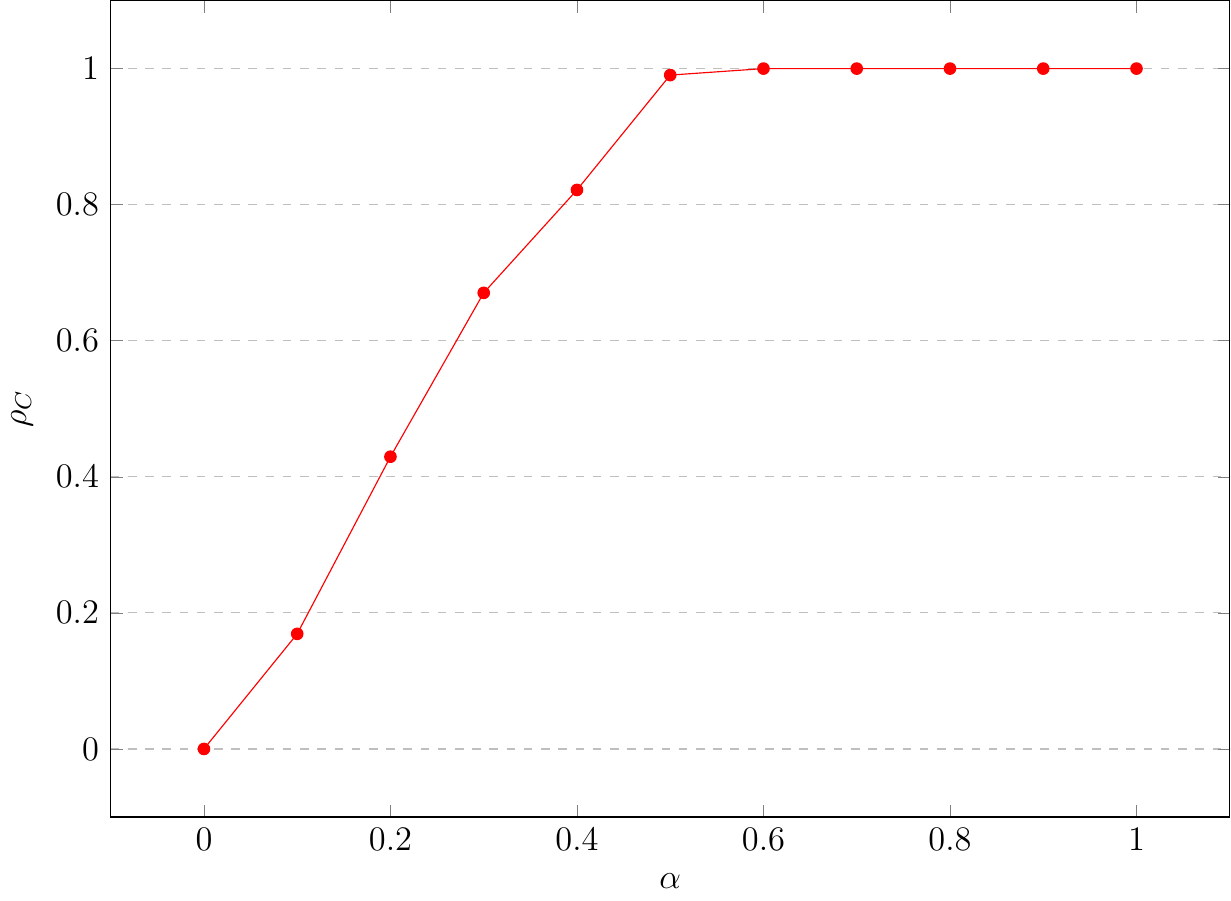}\\
\caption{Cooperation level in dependence of $\alpha$  for bimodal distribution when $r/G=0.7$. By increasing $\alpha$ we practically increase the fraction of selective players and the maximum cooperation level can be reached at around the percolation threshold.}\label{bimodal_rG070}
\end{figure}

The key feature of our present work is to allow social diversity among individuals and consider its possible consequence on the whole population. In connection to the results summarized in Fig.~\ref{bimodal_1} we already noted that a smaller fraction of individuals are capable to alter the total outcome of the system. This effect is highlighted in Fig.~\ref{bimodal_rG070} where we plotted the cooperation level in dependence of $\alpha$ for bimodal distribution at a fixed value of synergy factor $r/G=0.7$. Notably, this synergy factor value does not allow cooperators to survive in the traditional model when $\alpha=0$. However, as we increase $\alpha$, hence the portion of the selective players, then the general cooperation level grows. Importantly, there is no need for everyone to be in selective mode in order to reach the maximum cooperation level.

\begin{figure}
\centering
\includegraphics[width=9.5cm]{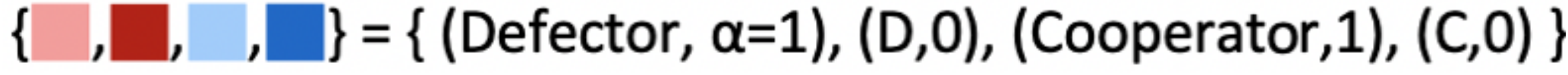}\\
\includegraphics[width=5.5cm]{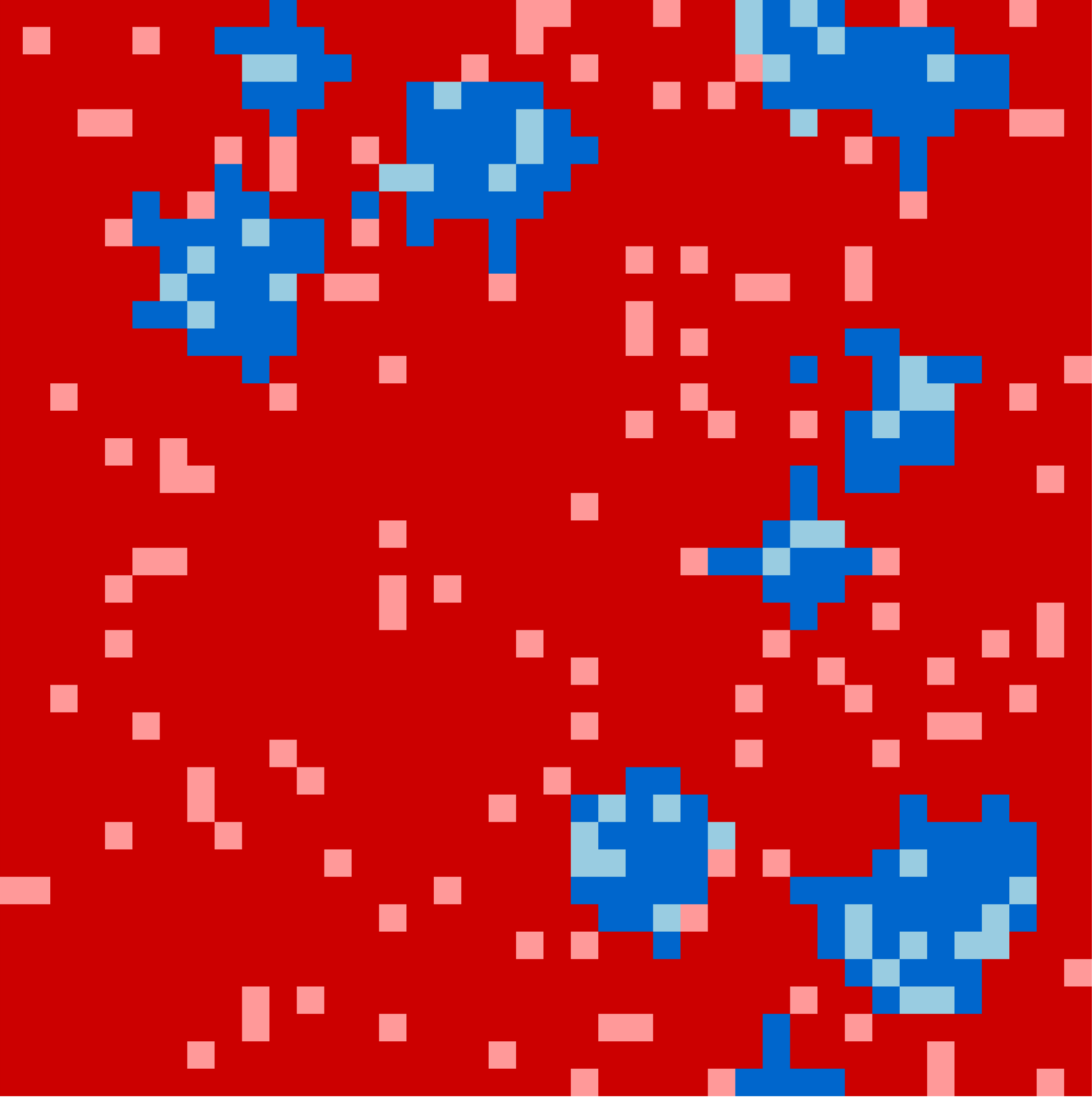} \hspace{1cm}
\includegraphics[width=5.5cm]{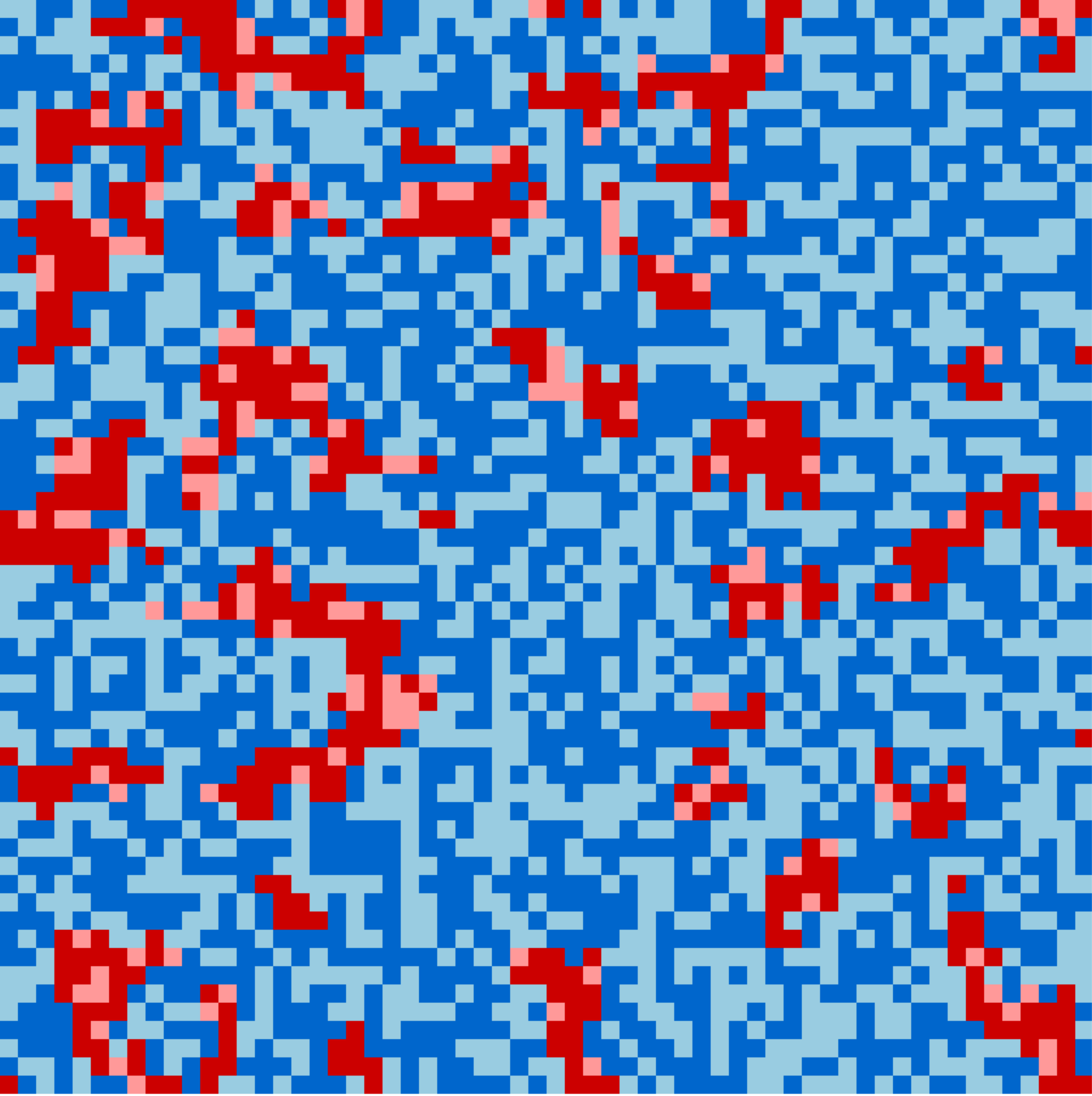}\\
\caption{Typical spatial distribution of strategies for bimodal distribution for $\alpha=0.1$ (left) and $\alpha=0.4$ (right). Defectors (cooperators) are marked by red (blue) colors where traditional (selective) players are denoted by dark (light) shades, as indicated by the legend. When selective players are rare (left) then cooperators can survive just in the vicinity of light blue selective cooperators. When their density is close to the percolation threshold (right panel) then they are not isolated anymore and can support the positive influences of each other to reach a high general cooperation level. Notably, in both cases $r/G=0.7$ would result in a full defection in the traditional model. The plotted system size is $L \times L = 100 \times 100$.}\label{alp}
\end{figure}

To get a deeper understanding of the described effect in Fig.~\ref{alp} we have plotted the spatial distribution of strategies at two $\alpha$ values. Here we used a specific coloring to mark not just the actual strategy of an individual but also its status describing how to invest in external groups. Accordingly, light blue (red) denotes a selective individual who is presently a cooperator (defector), while dark blue (red) denotes a traditional player in a cooperator (defector) state. The left panel shows a typical spatial distribution at a small $\alpha$ value where selective players are rare. Here the cooperative domains can only emerge around selective players who play the role of a seed of these domains. A traditional cooperator far from such a light blue seed cannot survive. But this problem can be resolved by increasing the density of selective players. This happens in the right panel of Fig.~\ref{alp} where the $\alpha$ value is close to the percolation threshold of the square lattice. As a result, the clouds surrounding light blue selective cooperators overlap, which ensures a safe environment for normal cooperators to survive. 

\section{Discussion}
\label{end}

In this paper, we proposed a protocol in a {\it PGG} setting that a cooperator could concentrate her external contribution to the game organized by the most successful neighbor. Different distributions of the selective players are examined and they all show the same conclusion, that is, given a fixed level of synergy factor, the final cooperation level raises when the amount of ``selective" players increases. In other words, a high level of selective players brings more cooperative behavior. We also investigated the spatial distribution of strategies for different ratios of these selective players. We found that when the level is close to the percolation threshold, the selective cooperators unite and create a supporting region of cooperation. 

The main message of our present study is that there is no need for everyone to possess some advanced skill and monitor the neighborhood how partners perform in their own venture. It is enough if just a portion of players has such a capacity and focus their investment on a specific venture which is functioning well. This directed flow of investment can strengthen network reciprocity locally and maintain a cooperator domain even at a harsh condition where traditional setup would result in full defection otherwise. If the mentioned selective players are distributed in space to cover their influence on the whole population then the system can switch the evolutionary trajectory completely and may terminate onto a full cooperator state.

It is worth stressing that the non-homogeneous investment protocol can help only if it targets the most successful group in the neighborhood. Put differently, if selective players reward one of their neighboring groups randomly, independently of its success, then we will obtain the outcome of the traditional model where network reciprocity works moderately. It simply means that supporting the more successful group is the guideline independently of the strategy of the player who organizes it. The latter is crucial because in this way we revealed a strategy-neutral mechanism which indirectly supports the well-being of the whole community. 

Last, some future explorations are valuable to consider. In this work, we used the von~Neumann-neighborhood where groups size $G=5$, which is the most frequent applied topology in the literature \cite{yang_lh_csf21,yang_hx_epl20,quan_jsm20}. However, it could be a straightforward extension to use Moore-neighborhood where we have larger groups $G=9$. Evidently, the latter is more demanding technically because we should check more neighbors and more groups when determine the group which is awarded by the enhanced contribution. We should also stress that
we only examined the suggested protocol on a lattice, while some heterogeneous topologies like scale-free networks or even adaptive networks are not explored yet. However, as we already argued above, the combination of heterogeneous topology and selected investment protocol will probably result in good conditions for cooperator strategy. To further increase the players' cognitive or learning ability, if a player could not only be selective on contributing her investment but also the players she interacts with, will an even higher cooperation level could be reached? Moreover, putting these selective players on different sites on different networks may also bring different levels of final cooperation level. 

\section*{Acknowledgement}
The research reported herein was partially supported by the Ministry of Science and Technology of the Republic of China (Taiwan), under grant No. 109-2410-H-001-006-MY3.

\bibliographystyle{elsarticle-num-names}

\end{document}